\documentclass[reprint,prl,superscriptaddress]{revtex4-1}
\usepackage{amsmath}
\usepackage[normalem]{ulem}
\usepackage{graphicx}
\usepackage{bm}
\usepackage{natbib}
\usepackage{color}

\def\mnras{Mon.Not.Roy.As.Soc.}

\def\ltsima{$\; \buildrel < \over \sim \;$}
\def\simlt{\lower.5ex\hbox{\ltsima}}
\def\gtsima{$\; \buildrel > \over \sim \;$}
\def\simgt{\lower.5ex\hbox{\gtsima}}
\graphicspath{{plots/}} 
 
\def\[{\begin{equation}}
\def\]{\end{equation}}

\renewcommand\section[1]{\emph{#1}.---}
\def\m@th{\mathsurround=0pt }
\def\eqalign#1{\null\,\vcenter{\openup1\jot \m@th
  \ialign{\strut\hfil$\displaystyle{##}$&$\displaystyle{{}##}$\hfil
  \crcr#1\crcr}}\,}

\begin{document}

\author{Fergus Simpson}
\email{fergus2@icc.ub.edu}
\affiliation{ICC, University of Barcelona (UB-IEEC), Marti i Franques 1, 08028, Barcelona, Spain.}
\author{Raul Jimenez}
 \email{raul.jimenez@icc.ub.edu}
 \affiliation{ICC, University of Barcelona (UB-IEEC), Marti i Franques 1, 08028, Barcelona, Spain.}
\affiliation{ ICREA, Pg. Lluis Companys 23, 08010 Barcelona, Spain.} 
\affiliation{Radcliffe Institute for Advanced Study, Harvard University, MA 02138, USA.}
\affiliation{Institute for Theory \& Computation, Harvard University, 60 Garden Street, Cambridge, MA 02138, USA.}
\author{Carlos Pena-Garay}
\email{penya@ific.uv.es}
\affiliation{Instituto de Fisica Corpuscular, CSIC-UVEG, P.O.  22085, Valencia, 46071, Spain.}
\affiliation{Laboratorio Subterr\'aneo de Canfranc, Estaci\'on de Canfranc, 22880, Spain.}
\author{Licia Verde}
\email{liciaverde@icc.ub.edu}
\affiliation{ICC, University of Barcelona (UB-IEEC), Marti i Franques 1, 08028, Barcelona, Spain.}
\affiliation{ ICREA, Pg. Lluis Companys 23, 08010 Barcelona, Spain.} 
\affiliation{Radcliffe Institute for Advanced Study, Harvard University, MA 02138, USA.}
\affiliation{Institute for Theory \& Computation, Harvard University, 60 Garden Street, Cambridge, MA 02138, USA.}
\affiliation{Institute of Theoretical Astrophysics, University of Oslo, Oslo 0315, Norway.}

\date{\today}
\newcommand{\ud}{\mathrm{d}}
\newcommand{\om}{\Omega_m}
\newcommand{\lcdm}{{$\Lambda$CDM}}
\newcommand{\xvec}{\vec{x}}
\newcommand{\vvec}{\vec{v}}
\newcommand{\avec}{\vec{a}}
\newcommand{\gnu}{\gamma_\nu}
\newcommand{\gnudot}{\dot{\gamma}_\nu}
\newcommand{\gnubar}{\bar{\gamma}_\nu}
\newcommand{\gnuddot}{\ddot{\gamma}_\nu}
\newcommand{\ubar}{\bar{u}}
\newcommand{\phidot}{\dot{\phi}}
\newcommand{\phiddot}{\ddot{\phi}}
\newcommand{\nneu}{n_\nu}
\newcommand{\co}{P}

\newcommand{\qt}{q}

\newcommand{\mH}{\mathcal{H}}

\newcommand{\pot}{U}
\newcommand{\uprime}{ v \frac{\partial \pot}{\partial x}}
\newcommand{\phiprime}{ \vec{v} \cdot \frac{\partial \Phi}{\partial \vec{x}}}
\newcommand{\veff}{V_{\text{eff}}}

\newcommand{\mnu}{m_\nu}
\newcommand{\dnu}{\delta_\nu}
\newcommand{\cnu}{c_{s\nu}^2}
\newcommand{\tnu}{\theta_\nu}

\title{Dark energy from the motions of neutrinos}

\begin{abstract}   
We demonstrate that a scalar field is unable to reverse its direction of motion while continuously exchanging energy with another fluid. If the rate of transfer is modulated by the scalar's acceleration, the field can undergo a rapid process of freezing, despite being displaced from the local minimum of its potential.  This enables dark energy to form from any potential, regardless of its shape. The field's equation of state mimicks that of a cosmological constant. We present a physically motivated realisation in the form of a derivative neutrino-majoron coupling.  Coherent motions, which form only once the neutrinos become non-relativistic, could be responsible for instigating the freezing process. This would provide a natural resolution to the dark energy coincidence problem, while avoiding the dynamical instabilities associated with mass-varying neutrino models. Finally we discuss possible means by which this model could be experimentally verified. 
\end{abstract}

\maketitle

\section{Introduction}

\begin{table*}
\begin{tabular}{ |l|c|l| } 
 \hline
 \bf{Event}   	& \bf{Scale Factor}   &  \bf{ Correlations} \\ 
 \hline
 1) Matter-Radiation Equality   & $10^{-3.5}$ &  \\ 
 2) Recombination   & $10^{-3}$ &  \\ 
 3) Neutrinos become non-relativistic   & $10^{-2}$ &  \\ 
  4)  Dark Energy - Matter Equality   & $1$ &  \\ 
\hline
 5) Neutrino-Radiation Equality &  $10^{-2}$  & Caused by (3)  \\ 
 6)  Nonlinear Density Perturbations &  $10^{-1}$ & Requires (1) \\ 
   7) Our Existence  & 1 & Requires (6)  \\ 
 \hline
\end{tabular}
\caption{A selection of recent cosmic events.  A minimum of three distinct events is required in order for the four fluids to reverse their positions in the density hierarchy. These are represented here by three moments of equality. Known correlations limit the number of independent events in this table to four. As-yet undiscovered correlations may further reduce this number.}
\label{tab:timeline}
\end{table*}

 %
 %
In the standard Lambda Cold Dark Matter (\lcdm) paradigm, each of the four key components (photons, neutrinos, dark matter, dark energy) crossed paths within the past ten e-folds. This corresponds to seven points of equality, in addition to other notable events such as recombination and the onset of nonlinear structure formation.  

The traditional coincidence problem in cosmology, casually stated as ``why now?", relates to the close proximity in scale factor between our existence and the onset of dark energy domination. This is arguably the most straightforward coincidence to resolve - we are better described as a sample in time rather than a sample in scale factor. Therefore if the Universe were to recollapse or rip within the next trillion years, our existence is not especially close to the onset of dark energy. Some of the other coincidences appear more stubborn, however. Why did the four major contributors to the cosmic energy budget all cross paths with each other in rapid succession?

The seemingly congested cosmic timeline, as summarised in Table \ref{tab:timeline}, can be remedied if some events are correlated with each other. For example, we should not be surprised to find that nonlinear structure formation initiated within a few e-folds of our existence, since it is a prerequisite for complex life. Furthermore, nonlinear structure formation was catalysed by the onset of matter domination, so these should not be interpreted as wholly independent events. Can we reduce this list of four independent events still further? In particular, could the onset of dark energy be a direct consequence of another recent event?  Upper bounds on the coupling strength between dark energy, dark matter, and neutrinos are extremely weak \citep{simpscat}. It is therefore feasible that non-gravitational interactions were responsible for the onset of cosmic acceleration.  

A number of models have been proposed which invoke energy exchange between a scalar field and a second fluid \citep{2000PhRvD..62d3511A}.  Motivated by the approximate equivalence of the neutrino mass and the dark energy density, a direct connection between the two has already been investigated in some detail \citep{fardon2004dark, kaplan2004neutrino, 2006Brookfield}. However these models tend to suffer from instabilities in their perturbations \cite{2005Afshordi}.  Recently a class of models involving a derivative coupling at the fluid level were presented in Ref.~\cite{2013Pourtsidou}, and explored further by Refs.~\cite{2015Skordis, 2016Pourtsidou, 2016BaldiSimpson}.  

In this work we shall first explore the means by which energy exchange can facilitate the formation of dark energy.  Then we present the particular case of a derivative coupling at the particle level, between neutrinos and a scalar field.

 \begin{figure}[b]
\includegraphics[width=80mm]{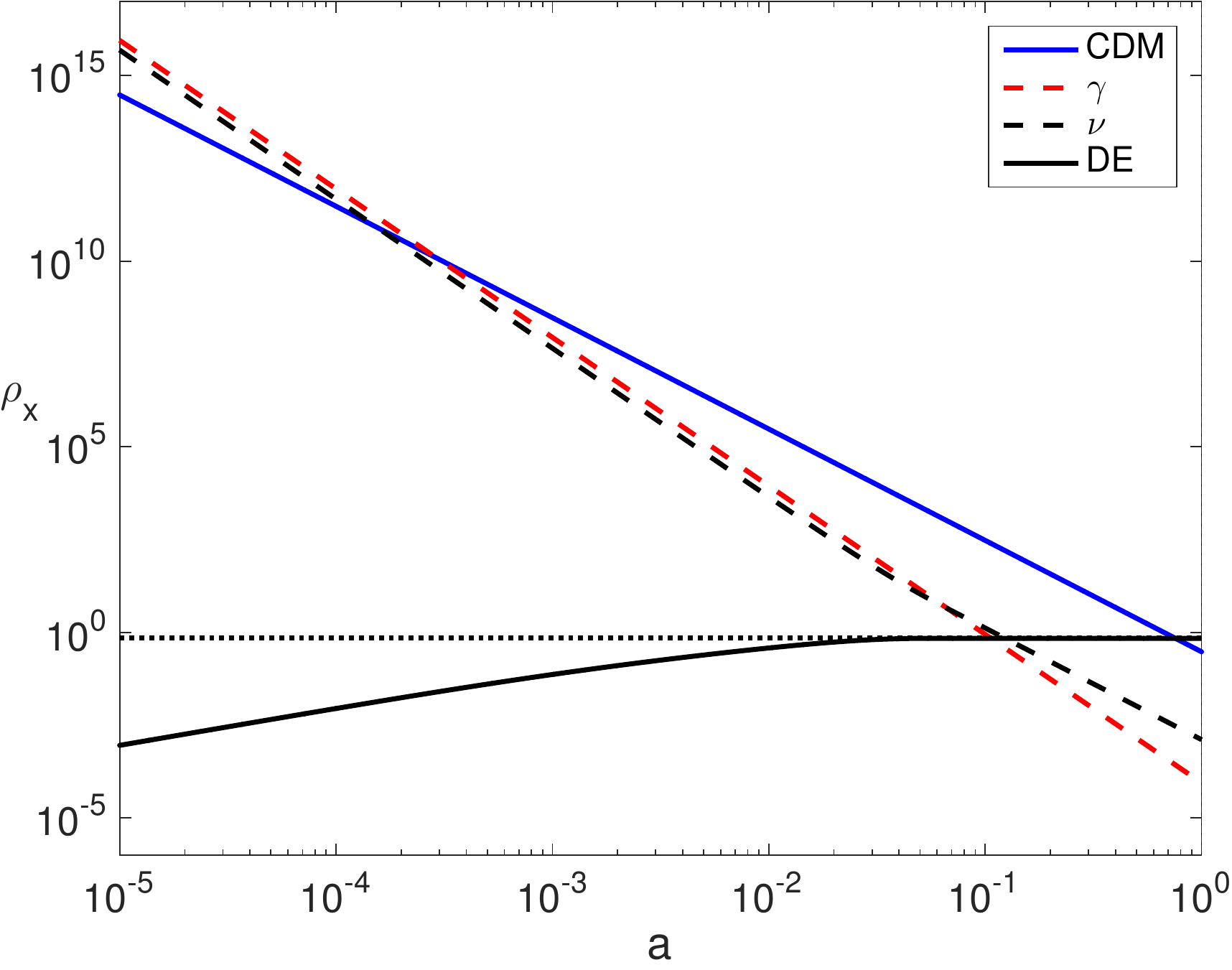}
\caption{ The energy densities associated with the four major cosmological fluids (photons, neutrinos, dark matter, and dark energy) in units of the present day critical density. The solid line represents a toy model which demonstrates how the dark energy could acquire its present day value, while having a negligible impact on the other fluids. 
 \label{fig:densities}}
\end{figure} 

\section{Freezing a Scalar Field}
In the presence of energy exchange, the modified Klein-Gordon equation is given by \cite{2008Bohmer, 2009PhRvD..79d3522C}  
\[ \label{eq:ants}
\ddot{\phi} +  3H \phidot  + V_\phi =  -\frac{Q}{\phidot} \, ,
\]
where $V_\phi \equiv \ud V / \ud \phi$ and the coupling term $Q$ dictates the rate at which the field loses energy to a secondary fluid.   While a number of theoretical models take the form  $Q \propto \phidot$, here we shall focus on those for which $Q =  \qt \phiddot$, where the function $\qt(t)$ may be time-dependent.  

In the regime where the scalar is moving sufficiently slowly, such that $|\phidot| \ll |\qt|$ and  $|\phidot| \ll | V_\phi  H^{-1} |$, then 
\[ \label{eq:phiddot}
 \ddot{\phi} = - \frac{V_{\phi}}{\qt} \phidot  \, .
\]
Instead of acting with its full `force', the influence of the potential $ V_\phi$ becomes progressively weaker as the scalar slows. The general solution for $\phi(t)$ is given by
\[
 \phi(t) = \phi(t_0) + \phidot(t_0) \int_{t_0}^t \exp \left( -  \int_{t_0}^{t'} \frac{V_\phi(t'')}{\qt(t'')} t'' \ud t''  \right) \ud t'   \, .
\]
The integrand remains positive for all $t$, ensuring $\phi(t)$ is monotonic. 
Provided the damping timescale is faster than the evolution of $V_\phi/\qt$, the decay of the kinetic term may be approximated as
\[  \label{eq:phidot}
\phidot  \propto \exp{\left( - \frac{V_\phi }{\qt} t \right) }  \, .
\]
Despite the presence of a gradient in the potential $ V_\phi$,  the scalar is prevented from rolling backwards due to the presence of the coupling term. This holds for any energy transfer $Q$ that does not vanish as the field comes to rest. In other words, while any positive energy transfer persists, the field is never allowed to arrive at a complete stop. This is   apparent from the right hand side of Eq.~(\ref{eq:ants}), which would be pathological if we ever reached $\phidot = 0$. Once frozen in place, the fluid's energy budget is then dominated by $V(\phi)$, for any $V(\phi)>0$. The trajectory of (\ref{eq:phidot}) implies that the field's equation of state $w \equiv p/\rho$ rapidly converges towards that of a cosmological constant,
\[
w(t) \simeq - 1 + \left[1 + w(t_0) \right]  \exp{\left(-\frac{ 2  V_\phi }{\qt }  [t - t_0]  \right) }  \, .
\]

\section{Direction of Transfer}
There are two qualitatively distinct phenomenologies, as determined by the sign of the energy transfer $\qt$. If $\qt$ is positive, the field can become frozen as it is climbing uphill, and this state can be maintained with a vanishingly small injection of energy  $Q$.  Conversely if $\qt$ is negative, the field can freeze as it is rolling downhill, as some kinetic energy is  extracted from the scalar. 

Certain forms of $V(\phi)$, such as an inverse power law, have the attractive feature of tracking the background density \citep{steinhardt1999cosmological, sahlen2007quintessence}. This alleviates the fine-tuning requirements of the initial conditions. However in their simplest form, these tracking solutions struggle to reproduce an equation of state consistent with cosmological observations, $w\simlt-0.8$. If however some energy is \emph{extracted} from the field, then its downhill roll will grind to a halt.  Much like the uphill case, the fluid which is on the receiving end of the energy transfer need only intake a vanishingly small energy flux in order to permanently impede the scalar's descent.

\section{Available Energy Sources}
What are the viable sources or sinks associated with the energy transfer $Q$? Fig.~\ref{fig:densities} illustrates the entangled trajectories of energy densities from the four major cosmological components within the conventional $\Lambda$CDM mode. Within the neutrino and dark matter densities lie distinct contributions from their kinetic motions, as opposed to their rest masses.  While the present day kinetic energy density of neutrinos is several orders of magnitude lower than the dark energy density, one need only look back three e-folds before they reach a comparable magnitude. Furthermore this epoch coincides with a qualitative transition in behaviour for neutrinos of mass $m_\nu \sim 0.06 \, \text{eV}$, as this is when they become non-relativistic. For these reasons we shall focus on assessing the feasibility of instigating dark energy from the kinetic motions of neutrinos, although dark matter remains a viable candidate, especially as its kinetic motion at matter-radiation equality is comparable to the present day dark energy density. 

\section{Gradient Coupling}
In the Standard Model of particle physics, the only anomaly-free global symmetry is the difference in baryon number B and lepton number L.
This fact has led in the past to proposed extensions of the Standard Model which invoke a continuous symmetry B-L \cite{Chikashige:1980ui}. 
Its spontaneous breaking would give rise to a massless Goldstone boson called the majoron  \cite{Aulakh1982yn}, assumed to have a singlet nature which reduces its coupling to the Z boson, and modulate the cooling rate of red giants \cite{Fukugita:1982ep}.
At low energies, this model leads to a derivative coupling \cite{burgess1994new}, of the form
  
\[ \label{eq:coupling}
\eqalign{
\mathcal{L}_{\text{eff}} &=  g\,{\bar\nu \gamma_{\mu} \nu} \, \partial^{\mu} \phi  \, , \cr
                                    &=  g\,{j_{\mu}} \partial^{\mu} \phi  \, ,
}
\]
where the boson $\phi$ couples derivatively to the neutrino current $j_{\mu}$, defined by the the neutrino field ($\nu$), its adjoint ($\bar\nu$), and the Dirac matrices $\gamma_{\mu}$. 
An empirical constraint on the coupling constant $g$ is that it ought to be smaller than $10^{-6}/$M  \cite{Gando2012pj}, where M is the Majorana mass. This ensures we satisfy supernova cooling, solar neutrino, kaon decay and double-beta decay bounds. 
In this proposal, the scalar is truly massless, but carries lepton number charge (majoron). This scheme suggests the existence of heavy isosinglet neutrinos in the mass range of several hundred MeV, with significant mixing to the electron neutrino. These heavy neutrinos could manifest themselves in the decays (including neutrinoless double beta decay) or by non-universality in the weak interactions of electrons versus other leptons. Later we shall discuss possible experimental signals of this model.  

\section{Influence on the Neutrino}
To begin, we consider the equation of motion for a single neutrino in the presence of a derivative coupling. The Lagrangian becomes
\[ \label{eq:lagrangian}
\mathcal{L}_\nu = - m_\nu \sqrt{1 - v^2} + F(\co)  - \pot(\xvec)  \, ,
\]  
where $\co \equiv  u^\mu  \phi_\mu$,  while $U(\xvec)$ represents an external potential, and the function $F(\co)$ determines the energy associated with the derivative coupling.
The Euler-Lagrange equations reveal the particle's motion to be
\[   
\frac{\ud \gnu}{\ud t} = - \frac{\gnu F_{PP} \dot{\phi} \ddot{\phi} + \uprime }{ m_\nu + F_{PP} \dot{\phi}^2 } \, ,
\]
where $F_{PP} \equiv \ud^2 F / \ud \co^2$, $\gnu$ denotes the Lorentz factor, and $\phidot \equiv \ud \phi / \ud t$.  Note that the particle's trajectory is only perturbed from the canonical case in the presence of a second derivative in $F(P)$.

The total rate of energy gain per particle $Q_i \equiv   \ud E_\nu /\ud t $ (and therefore the energy extracted from the scalar) is governed by the evolution of the kinematic and coupling terms

\[
Q_i  = - \frac{ \gnu m_\nu  F_{PP} \dot{\phi} \ddot{\phi} + m_\nu \uprime}{ \mnu + F_{PP} \dot{\phi}^2 } + F_P \left( \gnudot \dot{\phi} + \gnu \ddot{\phi} \right)  \, .
\]
Hereafter we shall  consider the linear case $F(P) = g P$. In the absence of an external potential, we are   left with
\[
 Q_i = g  \gnu \ddot{\phi}     \, .
\]
The energy associated with the coupling is modulated by the scalar's acceleration.  The coupling constant $g$ may be too small to exert a significant influence on any given neutrino. However as is apparent from Fig.~\ref{fig:densities}, only a small proportion of energy needs to be extracted from the neutrino fluid in order to account for the present day dark energy density. 

\section{Influence on the Scalar Field}
The scalar field's behaviour is modified by the sum of contributions from the large number of cosmic neutrinos. The aggregate energy transfer is then given by  $Q = \nneu Q_i $, where  $\nneu$ denotes the number density of neutrinos.   A microphysical interaction specified by (\ref{eq:coupling}) translates to a Lagrangian density at the fluid level given by
\[ \label{eq:derivL}
S_\phi = - \int \ud^4 x \sqrt{-g} \left[\frac{1}{2} \nabla_\mu \phi  \nabla^\mu \phi  + V(\phi) + \nneu g \bar{P}
 \right]  \, ,
\]
where   the particle-averaged coupling is given by $\bar{\co} \simeq \gnubar \dot{\phi}$, where $\gnubar$ denotes the mean Lorentz factor of the neutrinos.   In this model the scalar undergoes freezing once $| \phidot | \ll   | g \nneu \gnu |$.

\section{Gravitational Potential}
Next we consider the additional influence of neutrinos moving in the presence of a gravitational potential $\Phi$. The scalar now experiences an effective potential
\[
\ddot{\phi} \left( 1 +  \frac{g \nneu \gnubar}{\phidot}    \right)  +  3H \phidot +   V_\phi  + g  m_\nu n_\nu  \left\langle \phiprime \right\rangle  = 0 \, .
\]
While the neutrinos remain relativistic, the ensemble averaged $ \left\langle \phiprime \right\rangle$ vanishes due to their isotropic free-streaming motion, although there may be some contribution from modes as they cross the horizon. However once they become non-relativistic, neutrinos coherently infall towards overdense regions. Their velocity vectors begin to align with the gradient of the potential, thereby activating the coupling term. This mechanism suggests a direct correlation between events (3) and (4) in Table \ref{tab:timeline}, thereby addressing the dark energy coincidence problem.

Scalar fields undergoing uphill trajectories at late times are often overlooked, however they remain a feasible scenario here, due to the scalar receiving a recent kick from the  bulk motions of neutrinos. This process could accelerate the field up the potential $V(\phi)$ until the gradient satisfies $V_\phi \simeq - g m_\nu n_\nu  \langle \phiprime \rangle $. Then as the field decelerates, the freezing mechanism comes into play, and prevents the field from rolling back down the potential. 

If the field is rolling downhill, then the neutrinos' bulk motion can act as a brake rather than a driving force. In our model, the brake is applied once neutrinos become non-relativistic, naturally setting an appropriate energy scale.  Once frozen, only a few e-folds of cosmic expansion are required before the field  begins to dominate the energy budget, ushering in a period of cosmic acceleration. 

\section{Stability of Perturbations} \label{sec:stability}
One of the key problems with models of mass-varying-neutrinos is the tendency for instabilities to develop within the neutrino density field \cite{2005Afshordi}. This can be attributed to a combination of two factors: (a) an imaginary adiabatic speed of sound $c_s$ and (b) a short coherence length $m^{-1}_\phi < H^{-1}$ . In general the adiabatic sound speed for a fluid is given by
\[
c_s^2 = w - \frac{\dot{w}}{3H(1+w)} \, .
\] 
Despite often possessing a negative $c_s^2$, minimally coupled scalar fields are able to maintain stable perturbations  due to their low effective mass. On the other hand, standard mass-varying-neutrino models generically possess a large effective mass, and therefore a small coherence length $m^{-1}_\phi \ll H^{-1}$ \citep{2005Afshordi}. Their sub-horizon perturbations are therefore vulnerable to instabilities.  

In the case of derivative coupling, $\phi$ makes no explicit contribution to the effective potential. For example if we take $V(\phi) \sim \phi^4$, and using $V(\phi) \simlt H^2$, then we find $m^{-1}_\phi \simgt H^{-1}$. Therefore in this example,  sub-horizon perturbations never enter the adiabatic regime. 

Another source of instability can arise due to inhomogeneities the energy transfer itself \cite{valvi}. This is predominantly associated with models in which the dark matter density exclusively controls the rate of energy transfer.  If the rate of energy transfer is regulated by the dark energy fluid, as it is in our case, then stability can be achieved \cite{Jackson, 2009Gavela}. 

\section{Experimental predictions} \label{sec:experiments}
Earlier we advocated a neutrino-boson derivative coupling, where the boson may or may not carry a U(1) charge of lepton number (neutral or charged majoron).  Here we shall discuss some consequences of each scenario,
and their possible signatures in laboratory experiments, both in decays (including neutrinoless double beta decay) and non-universality in the weak interactions of electrons versus other leptons.

The observation of neutrinos from SN1987A and their temporal distribution indicates that majoron emission does not play a dominant role in core collapse processes. This leads to exclusions of the effective majoron-electron neutrino coupling constant,  g$_{ee} < 4 \cdot10^{-7} $  for the ordinary majoron-emitting decay mode  \cite{Gando2012pj}. Future double beta decay experiments which attain a sensitivity to the effective neutrinoless double beta mass m$_{\beta \beta}$  better than 20 meV (normal hierarchy neutrino splittings) may be capable of validating our model.

Majoron models can avoid strong constraints by carrying lepton number L$=-2$. Some features of a charged majoron model are: (a) the electron sum-energy spectrum in double-beta decay can be used to distinguish the two classes of models; (b) the decay rate for the new models depends on different nuclear matrix elements than ordinary majorons; and (c) all models require a (pseudo) Dirac neutrino, having a mass of a several hundred MeV, which mixes with electron neutrinos. In this case, the effective majoron-electron neutrino coupling constant is weakly constrained, g$_{ee} < 0.01$, with room for progress by the currently operational and future neutrinoless double beta decay experiments.  

\section{Conclusions}
Conventional models of quintessence generally suffer from the same kind of fine-tuning problems as the cosmological constant they were designed to usurp. Ordinarily, the scalar must either lie at a local minimum of a finely-tuned potential, or be at a finely-tuned location within an extremely flat potential. However we have demonstrated that if a scalar field is in the process of energy exchange, it can mimic a cosmological constant even when displaced from the local minimum of its potential. This allows potentials of \emph{any} shape to act like a cosmological constant. 

The underlying phenomenology of this freezing process is founded in classical mechanics. A cyclist who is pedalling uphill might begin to struggle against an increasingly steep gradient. Yet no matter how feeble the rider's power output becomes, they can always maintain a constant pedalling rate, simply by selecting an appropriately high gear. They can \emph{never} roll backwards.

We have demonstrated that a derivative coupling between neutrinos and the majoron could provide the required energy transfer. As neutrinos become non-relativistic, their velocities begin to align with the gradients in the local gravitational potential. In the presence of a derivative coupling, this creates an effective potential for the scalar field. The effective potential can act as either the driving force or a brake, depending on the sign of the coupling constant.  In each case, once the freezing process sets in, the field's equation of state rapidly approaches that of a cosmological constant. 

One path towards resolving these issues may be anthropic selection from within an ensemble: a multiverse. This has been touted as an explanation for the value of the cosmological constant \citep{1987PhRvL..59.2607W, 1995MNRAS.274L..73E, 2007PeacockAnthropic, 2015Piran} and the neutrino mass \cite{2005Tegmark, 2015Bousso}. However, before accepting a hypothesis which is not experimentally falsifiable, it is wise to exclude all viable alternatives which \emph{are}.

It has been argued in the past \cite{2001Witten} that any theory of quantum gravity will not allow B-L to be a symmetry of nature and thus could provide neutrinos with their mass. It is, therefore, curious that dark energy could indirectly result from broken symmetries, due to the quantum nature of gravity; the so-called IR-UV connection.

In summary, the freezing mechanism presented in this work offers a  generic means to generate dark energy from any scalar potential $V(\phi)$.  The derivative coupling model we propose offers a natural and experimentally verifiable solution to the dark energy coincidence problem.

\begin{acknowledgments}
We thank Alan Heavens, Antony Lewis, John Peacock, Nuria Rius and David Spergel for useful insights and comments.
We acknowledge support by Spanish Mineco grant AYA2014-58747-P and  MDM-2014-0369 of ICCUB (Unidad de Excelencia `Mar\'ia de Maeztu').
\end{acknowledgments}



%

\end{document}